\documentclass[prb,twocolumn,superscriptaddress,longbibliography,notitlepage]{revtex4-2}
\usepackage{amssymb}
\usepackage{amsmath}
\usepackage{graphicx}
\usepackage{hyperref}
\usepackage{xcolor}

\newcommand{\comment}[1]{}

\begin{document}

\title{Hybrid convolutional neural network and PEPS wave functions for quantum many-particle states}

\author{Xiao Liang}
\affiliation{Institute for Advanced Study, Tsinghua University, Beijing, 100084, China}

\author{Shao-Jun Dong}
\affiliation{CAS Key Laboratory of Quantum Information, University of Science and Technology of China, Hefei 230026, Anhui, China}
\affiliation{Synergetic Innovation Center of Quantum Information and Quantum Physics, University of Science and Technology of China, Hefei, 230026, China}
\author{Lixin He}
\email{helx@ustc.edu.cn}
\affiliation{CAS Key Laboratory of Quantum Information, University of Science and Technology of China, Hefei 230026, Anhui, China}
\affiliation{Synergetic Innovation Center of Quantum Information and Quantum Physics, University of Science and Technology of China, Hefei, 230026, China}

\begin{abstract}

Neural networks have been used as variational wave functions for quantum many-particle problems. It has been
shown that the correct sign structure is crucial to obtain the high accurate ground state energies.
In this work, we propose a hybrid wave function combining the convolutional neural network (CNN) and projected entangled pair states  (PEPS), in which the sign structures are determined by the PEPS, and the amplitudes of the wave functions are provided by CNN. We benchmark the ansatz on the highly frustrated spin-1/2 $J_1$-$J_2$ model.  We show that the achieved ground energies are competitive to state-of-the-art results.

\end{abstract}
\maketitle

\section{introduction}
Neural network (NN) based machine learning has been applied to solve
various physical problems \cite{review1,review2}, such as experiment automation\cite{143km}, quantum state classification\cite{leiwang,phase1,phase3,topo1,topo2}, emerging physics from neural networks\cite{emerge1}, simulation of quantum computation\cite{QC1}, accelerating monte-carlo calculations\cite{MC1,MC2,MC3}, accelerating density-function-theory calculations\cite{DFT1,DFT2}, representing quantum states\cite{represent1,represent2,represent3,represent4,ansatz1,ansatz3,ansatz5,ansatz7,ansatz8,rbm_science} and time evolution for open quantum systems\cite{open1,open2,open3,open4}.

Recently, neural networks have also been applied to solve the quantum many-particle problems,
which is one of the most interesting and challenging fields in condensed matter physics.
A variational ansatz, namely restricted Boltzmann machine (RBM)
has been demonstrated that can solve the non-frustrated Heisenberg model
to a high accuracy\cite{rbm_science}, that is comparable to the state of art methods.
It has been argued that RBM \cite{represent1,represent3}
and convolutional neural network (CNN) \cite{CNNvolumelaw} can even
represent quantum states beyond area law entanglement, and therefore
have great potential to solve a large class of quantum many-particle problems.

Solving the quantum frustrated models is an even more challenging problem for neural networks.
Some of the authors first attacked the highly frustrated spin-1/2 $J_1$-$J_2$ model on the square lattice
via the convolutional neural networks (CNN).\cite{CNNj1j2_1}
They have obtained the ground state energies that is lower than
the string-bond-states method\cite{SBS}.
Later on, Choo et. al. found that the ground energies can be significantly improved
by introducing a prior sign structure before CNN and enforcing the rotational symmetry\cite{CNNj1j2_2}.
However, in the work, the sign structures are artificially assigned based on the Marshall-Peierls sign rule (MPSR)\cite{MPSR}.
Westerhout et. al.  investigate the learning ability for the sign structure
and amplitudes of the wave functions of quantum frustrated systems, using supervised ML.
It was concluded that while CNNs have no problem to generate the amplitudes
of the wave function, the generation of sign structure is very challenging, especially for
the frustrated systems\cite{CNN_frustration}.
Since the sign structure is a discontinues function with respect to spin configurations, it is difficult
to present the sign structure and the amplitude of the wave function by a single NN.
Szab\'o et. al. introduced a CNN structure for the quantum many-particle wave functions,
in which the sign structure is represented by a single-layer CNN, and the amplitudes of the wave function are
represented by a separate deep CNN. \cite{CNNj1j2_3}.

On the other hand, the tensor network methods, e.g., the projected-entangled-pair-states(PEPS) method,\cite{PEPS_1,PEPS_2,PEPS_4,PEPS_7,PEPS_8} has achieved extremely
high precision in solving the ground energies of quantum many-particle problems, especially for highly frustrated systems\cite{PEPS_10,PEPS_11,PEPS_12,Kagnome,corboz,tj}.
In principle, the PEPS can present any quantum state faithfully, provided the tensor bond dimension is large enough.
However, the computational scaling is extremely high to contract a PEPS, especially for a PEPS with periodic boundary conditions (PBC), which limits their applications.

In this work, we propose a hybrid PEPS and deep CNN hybrid structure, which combines the two state-of-art techniques,
as a variational ansatz for the quantum many-particle problems.
For a given spin configuration, the corresponding coefficient is the product of the sign
(+ or -) provided by PEPS (with small bond dimension) and the magnitude provided by CNN.
We first show that he MPSR can be rigorously represented by the PEPS even with virtual bond dimension $D$=1
on the square lattice. More complicated sign rules should be able to be represented by PEPS with a larger bond dimension.
Another advantage of using PEPS to present the sign structure is that the PEPS can be optimized via
time-evolution method without sampling over the spin configurations,
which may bypass the spin generation problem.\cite{CNN_frustration}
We benchmark the ansatz by calculating the spin-1/2 $J_1$-$J_2$ model on the square lattice.
We show that the sign rule provided by the PEPS can significantly improve the ground state energies of the $J_1$-$J_2$ model, which are very competitive to other state-of-art neural network wave functions.
The results show that the PEPS+CNN hybrid structure is a very promising variational wave function for quantum many-particle problems.

%\section{deep cnn architecture}
\begin{figure}[t]
\includegraphics[width=1.0\columnwidth]{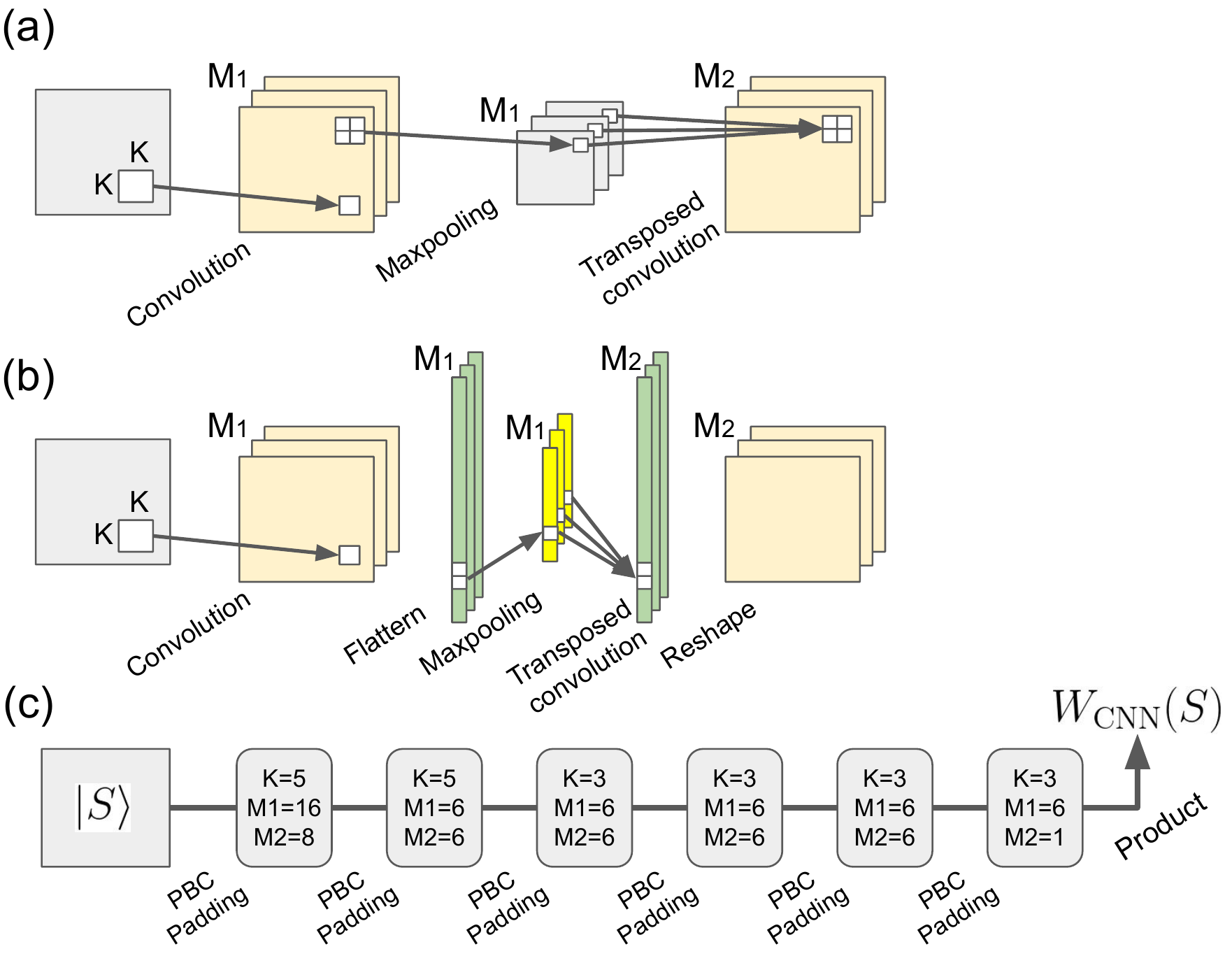}
\caption{ Illustration of the deep CNN, where $\mathrm{K}$ is the side length of the convolution filter, $\mathrm{M}_1$ and $\mathrm{M}_2$ are the output channel of convolution and transposed-convolution, respectively.
(a) The shallow CNN in Ref.\onlinecite{CNNj1j2_1};
(b) The building blocks for the deep CNN modified from the shallow CNN, in which the two dimensional
maxpooling and transposed-convolution are changed to one-dimensional;
(c)The deep CNN is built by stacking the building block (b) for six times. To maintain the dimension, the input of each block is padded based on PBC. The input of the deep CNN is a spin configuration on $|\sigma_1^z\sigma_2^z\cdots\sigma_N^z\rangle$ basis. The final output is denoted as $W_{\rm CNN}(S)$, it is generated by taking the products of the output neurons in the last block. }
\label{Fig:NN_structure}
\end{figure}

\section{Methods}

\subsection{Deep CNN architecture}

The building block of the deep CNN is based on the network used in a previous literature\cite{CNNj1j2_1} which is depicted in Fig.~\ref{Fig:NN_structure}(a). Each block consists of a convolution layer, a maxpooling layer and a transposed-convolution layer. In the figure, $K$ denotes the convolution filter size, $M_1$ and $M_2$ are the channel number of convolution and transposed-convolution respectively. To reveal the details of the CNN structure, we take a 4-sites spin-chain as an example. We use $M_1$=$M_2$=1 and $K$=3. The convolution layer performs the following transformation:
\begin{equation}
\begin{bmatrix}
p_1\\
p_2\\
p_3\\
p_4
\end{bmatrix}=
\bf{b}+
\begin{bmatrix}
w_1 &w_2  &w_3  &0  &0  &0 \\
0 &w_1  &w_2  &w_3  &0  &0 \\
0 &0  &w_1  &w_2  &w_3  &0 \\
0 &0  &0  &w_1  &w_2  &w_3
\end{bmatrix}
\begin{bmatrix}
s_4\\
s_1\\
s_2\\
s_3\\
s_4\\
s_1
\end{bmatrix},
	\label{cnndetail1}
\end{equation}
where $w$ are the weights of the filter and $\bf{b}$ is the bias vector, $[s_4,s_1,s_2,s_3,s_4,s_1]^\intercal$ is the input spin configurations with $s_i$=$\pm 1$. To maintain the dimension, each side of the input is padded $(K-1)/2$ sites according to PBC.

Then maxpooling is performed on the neurons,
\begin{equation}
\begin{bmatrix}
g_1\\
g_2\\
\end{bmatrix}=
\begin{bmatrix}
\rm{max}(p_1,p_2)\\
\rm{max}(p_3,p_4)\\
\end{bmatrix}.
	\label{cnndetail2}
\end{equation}
After the maxpooling, a transposed-convolution is performed,
\begin{equation}
\begin{bmatrix}
h_1\\
h_2\\
h_3\\
h_4
\end{bmatrix}=
\begin{bmatrix}
d_1 &0 \\
d_2 &0 \\
0 &d_1 \\
0 &d_2
\end{bmatrix}
\begin{bmatrix}
g_1\\
g_2\\
\end{bmatrix},
	\label{cnndetail3}
\end{equation}
to restore the original size of the lattice.

Each output channel after transposed convolution is the direct summation of the neurons from each filter\cite{CNNj1j2_1}:
\begin{equation}
\begin{bmatrix}
h_1\\
h_2\\
h_3\\
h_4
\end{bmatrix}=
\sum_{m=1}^M
\begin{bmatrix}
d_1^{(m)}g_1^{(m)}\\
d_2^{(m)}g_2^{(m)}\\
d_3^{(m)}g_3^{(m)}\\
d_4^{(m)}g_4^{(m)}
\end{bmatrix}.
	\label{cnndetail4}
\end{equation}

For a 2D spin lattice, the filters are 2D and the convolutions are performed in 2D. In the previous literature\cite{CNNj1j2_1}, the maxpooling and transposed-convolution is chosen naturally as two-dimensional, as  shown in Fig.~\ref{Fig:NN_structure}(a). However, because two-dimensional maxpooling neglects too many neurons, it reduces the representation ability  when the CNN is deep. To build a deep CNN structure, we modify the two-dimensional maxpooling and transposed-convolution to one-dimensional. The modified structure is shown in Fig.~\ref{Fig:NN_structure}(b). To fit the one-dimensional maxpooling, the output neurons of each convolution filter is also flattened to one-dimensional, and the output neurons after transposed-convolution are reshaped to maintain the dimension. Figure~\ref{Fig:NN_structure}(b) is the building block of our deep CNN. Each neuron after the transposed-convolution is a linear combination of the spin values:
\begin{equation}
	h_i=\sum_{k=1\cdots N}A_{i,k}\sigma_k^z+c_i,
	\label{network_output}
\end{equation}
where $A_{i,k}$ is the a real coefficient of spin $\sigma_i^z$ in $h_i$ and $c_i$ is a real number.

The deep CNN is built by stacking the building block six times, as shown schematically in Fig.~\ref{Fig:NN_structure}(c). The input of the first block is 2D spin configurations on the basis of $|S\rangle=|\sigma_1^z\sigma_2^z\cdots\sigma_N^z\rangle$, and the value of each spin $\sigma_i$ is $\pm 1$. To maintain the dimension, each channel of the input of each building block is padded $(K-1)/2$ sites according to PBC same as the one dimensional case. The output of the deep CNN is the wave-function coefficient $W_{\rm CNN}(S)$, which is the product of the output neurons in the final block. The wave-function that the deep CNN represents is:
\begin{equation}
	|\Psi_{\rm CNN}\rangle=\sum\limits_S W_{\rm CNN}(S)|S\rangle,
	\label{wavefunction}
\end{equation}
where $|S\rangle$ is the spin configuration. Because of the product of the output neurons from the last building block, the deep CNN associates the spin configuration to the high order correlations between the spins:
\begin{equation}
	W_{\rm CNN}(S)=\sum_{n_1\cdots n_N}g(n_1\cdots n_N;\sigma_1\cdots\sigma_N)\sigma_1^{n_1}\cdots\sigma_N^{n_N},
	\label{final_output}
\end{equation}
where the order number $\tilde{N}=n_1+n_2+\cdots+n_N\leq N$ and the coefficient $g$ is given by the deep CNN. Because of the maxpooling, $g$ also depends on the spin configurations.\cite{CNNj1j2_1}

The deep CNN structure used in this work has several important differences compared to the deep CNN structure used in Ref.~\onlinecite{CNNj1j2_2} and Ref.~\onlinecite{CNNj1j2_3}. First, the non-linearity in our deep CNN is induced by the maxpooling, whereas
in Ref.\onlinecite{CNNj1j2_2} and Ref.~\onlinecite{CNNj1j2_3}, the activation functions are used.
The maxpooling picks up the most important degree of freedom in a convolution filter, which is similar to the coarse-grained process in a renormalization group theory.\cite{pooling}

Another important difference is that, traditionally, the output wave functions are
taken as the exponential function of the NN.\cite{represent1,represent2,represent3,ansatz5,rbm_science},
whereas in our construction, the wave-function $W_{\rm CNN}(S)$ is generated by taking the products of the neurons from the last building block. \cite{CNNj1j2_1}
In our deep CNN, $W_{\rm CNN}(S)$ can be either positive or negative, which is crucial to represent the ground states of frustrated systems, using only real network parameters.

As discussed in Ref.\onlinecite{CNNj1j2_1}, the representability of the CNN structure relies on whether it can capture the long-range spin correlations (or entanglement). For the shallow CNN shown in Fig.~\ref{Fig:NN_structure}(a)(b), the first convolution layer is vital to capture the long-range correlation, therefore the CNN filter should be as large as the lattice size or the correlation length\cite{CNNj1j2_1}. However, in the deep CNN, the spins in different filters of front layers
can be entangled via maxpooling and convolution in the deeper neural networks,
therefore it can capture the long-range spin correlations efficiently
with much smaller filters. This is similar to the renormalization
process in the numerical renormalization group method.\cite{RG} The information is reused by stacking the building blocks,
which can enhance the representation ability\cite{CNNvolumelaw}.
In our investigations, we fix the deep CNN structure as depicted in Fig.~\ref{Fig:NN_structure}(c), there are 3531 real parameters. This is compared to the shallow CNN used in our previous work,\cite{CNNj1j2_1} where $M_1$=128 and $K=9$ filters were used, leading to 11009 real parameters. Obviously, the deep CNN has much fewer parameters.

%\section{PEPS}
\begin{figure}[t]
\includegraphics[width=1.0\columnwidth]{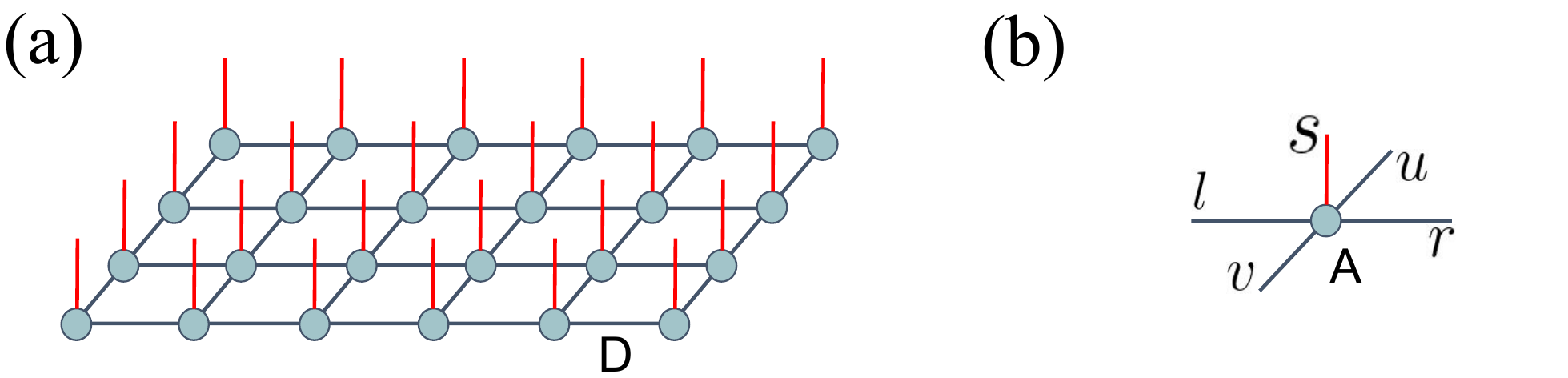}
\caption{(a) The illustration of a PEPS on a 4$\times$6 lattice. The circles stand for
a rank 5 tensor on the lattice sites, which are connected by the virtual bonds of dimension $D$;
(b) The rank 5 tensor $A(l,r,u,v,s)$ on each site has four virtual bonds: $l$, $r$, $u$ and $v$ and one physical index $s$.
}
\label{Fig:PEPS}
\end{figure}

When using a (deep) CNN to represent the quantum many-particle wave-functions, the wave-functions consist of sign structure and amplitudes. In many ML cases, especially for regression problems, the output is usually continuously distributed with respect to the continuous input.
However, this may not be true for a quantum system.
Taking the $J_1$-$J_2$ model as an example. It is known, that when $J_2$=0 the ground state exactly obeys the MPSR, that the sign of the wave function is $(-1)^{M_A}$ where $M_A$ is the magnetization of equivalent sublattice $A$. MPSR is a discontinuous function with respect to flipping spin configurations. Therefore, the sign rule is quite difficult
to be presented by a single CNN\cite{CNN_frustration} which must keep the amplitude smooth at the same time.\cite{CNNj1j2_3}
It has been demonstrated in Ref.~\onlinecite{CNNj1j2_2} that an explicit preconditioning to the CNN wave function
by a MPSR can significantly improve the ground state energies
compared to those without the preconditioning, for the $J_1$-$J_2$ model.
However, the sign rule is added {\it ad hoc} by hand in Ref.~\onlinecite{CNNj1j2_2}. Later on, Szab\'o et. al. introduce
a separate single-layer CNN to represent the sign structure, which is optimized
with fixed amplitudes. They obtained accurate MPSR for both $J_2$=0 and $J_2$=0.5$J_1$.\cite{CNNj1j2_3}

\subsection{PEPS-CNN hybrid structure}

We propose a hybrid PEPS and CNN structure as a variational wave function ansatz for quantum many-particle problem.
Considering a 2D square lattice with $N$=$L_x\times L_y$ sites, and
$d$-dimensional local Hilbert space (for a spin-1/2 system, $d$=2),
denoted as $|s_m\rangle$ on the site $m$=($i$,$j$),
the PEPS wave function of this system, which is schematically shown in Fig.~2(a),
can be written as\cite{PEPS_4},
%%
%\begin{equation}
%  |\Psi_{\rm PEPS}\rangle  = \sum_{s_m=1}^d { \rm Tr}  (A_{1}^{s_{1}} A_{2}^{s_{2}}
%  \cdots A_{N}^{s_{N}})  |s_{1} \cdots s_{N} \rangle,
%   \label{Eq:PEPS}
%\end{equation}
%%
\begin{equation}
  |\Psi_{\rm PEPS}\rangle  = \sum_S W_{\rm PEPS} (S) |S \rangle,
   \label{Eq:PEPS}
\end{equation}
where,
\begin{equation}
W_{\rm PEPS}(S)= {\rm Tr}  (A_{1}^{s_{1}} A_{2}^{s_{2}}
  \cdots A_{N}^{s_{N}}),
	\label{Eq:PEPS_WS}
\end{equation}
$A^{s_{m}}_{m}$=$A_{m}(l,r,u,v,s_{m})$ is a
rank-five tensor located on site $m$ as shown in Fig.~2(b).
The physical index $s_{m}$ takes value from $1$ to $d$
and four virtual indices $l,r,u,v$, which correspond to four nearest neighbors.
The dimension of each virtual bond is $D$. and
the ``${\rm Tr}$'' denotes the contraction over all the virtual indices of the tensor network.

One way to construct the PEPS-CNN hybrid structure is to take the direct products of the PEPS wave functions to the CNN wave functions. Here, we propose an alternative PEPS-CNN hybrid structure as follows,
\begin{equation}
	W(S)=\text{Sign}(W_\text{PEPS}(S))\cdot W_{\rm CNN}(S),
	\label{PEPS+CNN}
\end{equation}
%where $W_0$ denotes the output from the CNN
i.e., we take the sign of the PEPS wave function and multiply it to the CNN wave function.
We first show that the PEPS can rigorously represent the MPSR on the square lattice even with virtual bond dimension $D$=1. For both N\'{e}el order and stripe order, MPSR requires a spin-flip on the sublattice changing the wave function's sign. For $D$=1, the PEPS on each site ($i$,$j$) only has two elements ($a$, $b$), one for spin up and one for spin down. For the N\'{e}el order, we take $a$=1 for all sites and $b$=(-1)$^{i+j}$. For the stripe order, we take $a$=1 for all sites and $b$=(-1)$^j$. One can easily prove that these PEPSs satisfy MPSR, which have also been checked numerically.
One may expect that more complicated sign rules should be able to be presented by PEPS with larger $D$.

%Through our investigations,
The ground state many-particle wave functions respect
to the symmetry of the Hamiltonian. We could therefore also enforce the rotational symmetry to the wave function,\cite{C4_symmetry}
\begin{equation}
\tilde{W}(S)=\sum_{i=0}^3W(\hat{T}^iS),
	\label{C4}
\end{equation}
and $\hat{T}$ is the rotation operator that rotates the spin configuration for 90 degrees.

%\section{numerical investigations}
%\textit{Numerical investigations}--
\section{Results and discussion}

We benchmark the hybrid PEPS-CNN wave functions for the spin-1/2 $J_1$-$J_2$ model.
The Hamiltonian of the model reads,
\begin{equation}
	\hat{H}=J_1\sum_{\langle i,j \rangle}\textbf{s}_i\cdot\textbf{s}_j+
	J_2\sum_{\langle\langle i,j \rangle\rangle}\textbf{s}_i\cdot\textbf{s}_j,
	\label{eq:j1j2}
\end{equation}
where $\langle i,j \rangle$ and $\langle\langle i,j \rangle\rangle$ indicate the nearest and next-nearest neighbouring spins pairs. The model is calculated on a  $L\times L$ square lattice with PBC.
We set $J_1$=1 and $L$=10 throughout the investigations and on two cases: $J_2$=0 and $J_2$=0.5. When $J_2$=0.5, the frustration is very strong and the ground state is inferred as the quantum spin liquid\cite{QSL_1,QSL_2,QSL_3}, solving the ground state is challenging.

\begin{table}[]
\caption{Comparison between the ground energies of the spin-1/2 $J_1$-$J_2$ model on a
10$\times$10 square lattice,
achieved by different CNN structures.
S. CNN and D. CNN denote the shallow CNN and deep CNN respectively.}
\begin{tabular}{c|cccc}
\hline\hline
                & S. CNN& PEPS+S. CNN & PEPS+D. CNN &  \\ \cline{1-4}
$J_2$=0 & -0.668236   & -0.670718      & -0.671330    & \\
$J_2$=0.5 & -0.482986   & -0.492335        & -0.495502     & \\
\hline
\end{tabular}
\label{Tab:structure_comparison}
\end{table}
\begin{figure}[t]
\includegraphics[width=0.9\columnwidth]{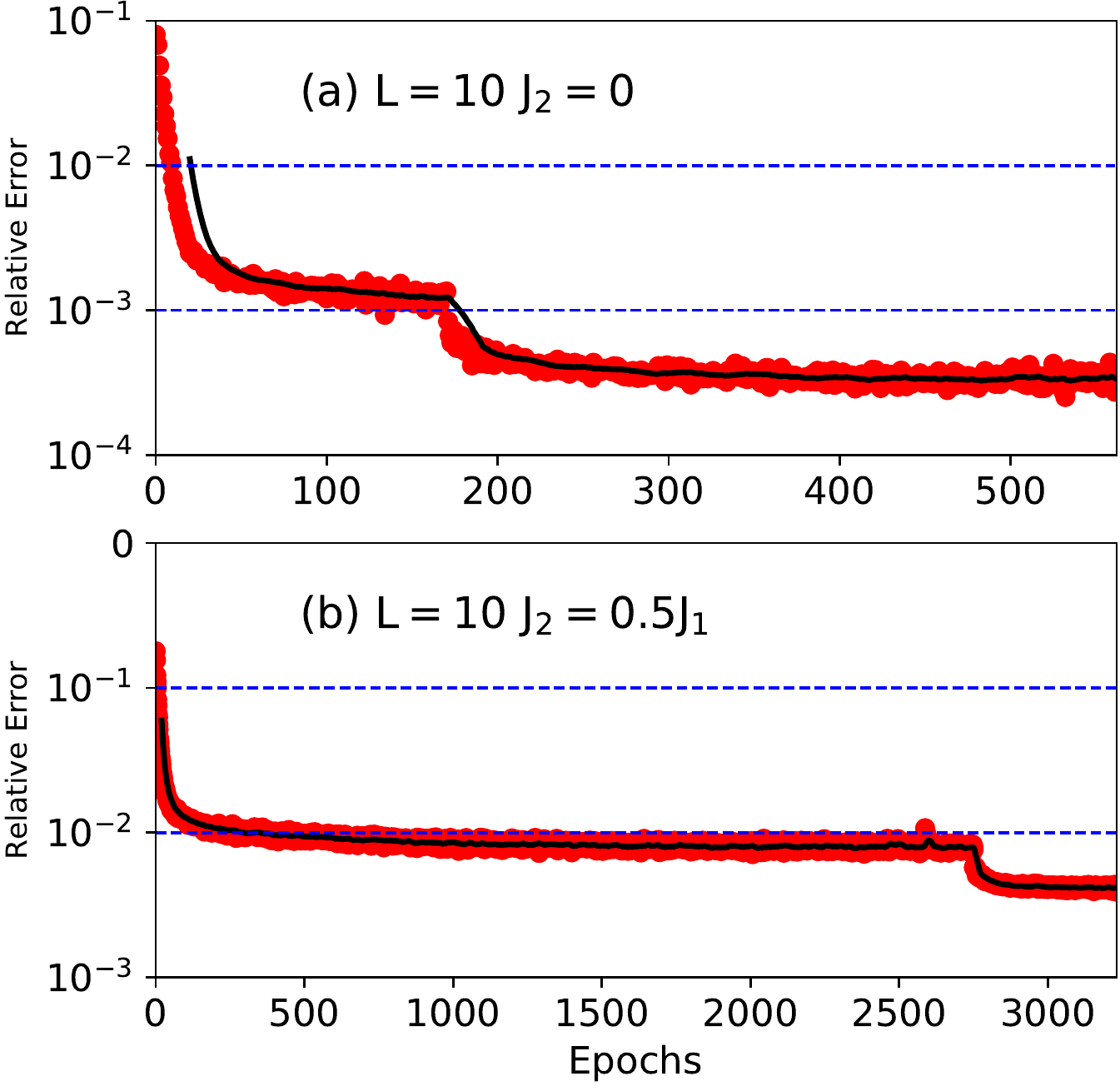}
\caption{ The optimization of the energy with respect to SR epochs for the PEPS+deep CNN,
in the case of (a) $J_2$=0 and (b)  $J_2$=0.5. The relative error is defined as $|(E-E_{\rm best})/E_{\rm best}|$,
where $E_{\rm best}$ are given in Table II.
The SR is firstly done without enforcing rotational symmetry, after energy converges, we enforce rotational symmetry on the wave-functions, which yields to a sharp decrease of the relative error. The sample number for each SR step is 16000 without rotational symmetry and 100000 with rotational symmetry. The energy expectation is averaged over the last ten SR steps, which is denoted by the solid black line.
}
\label{Fig:Precisions}
\end{figure}

\begin{figure}[t]
\includegraphics[width=0.8\columnwidth]{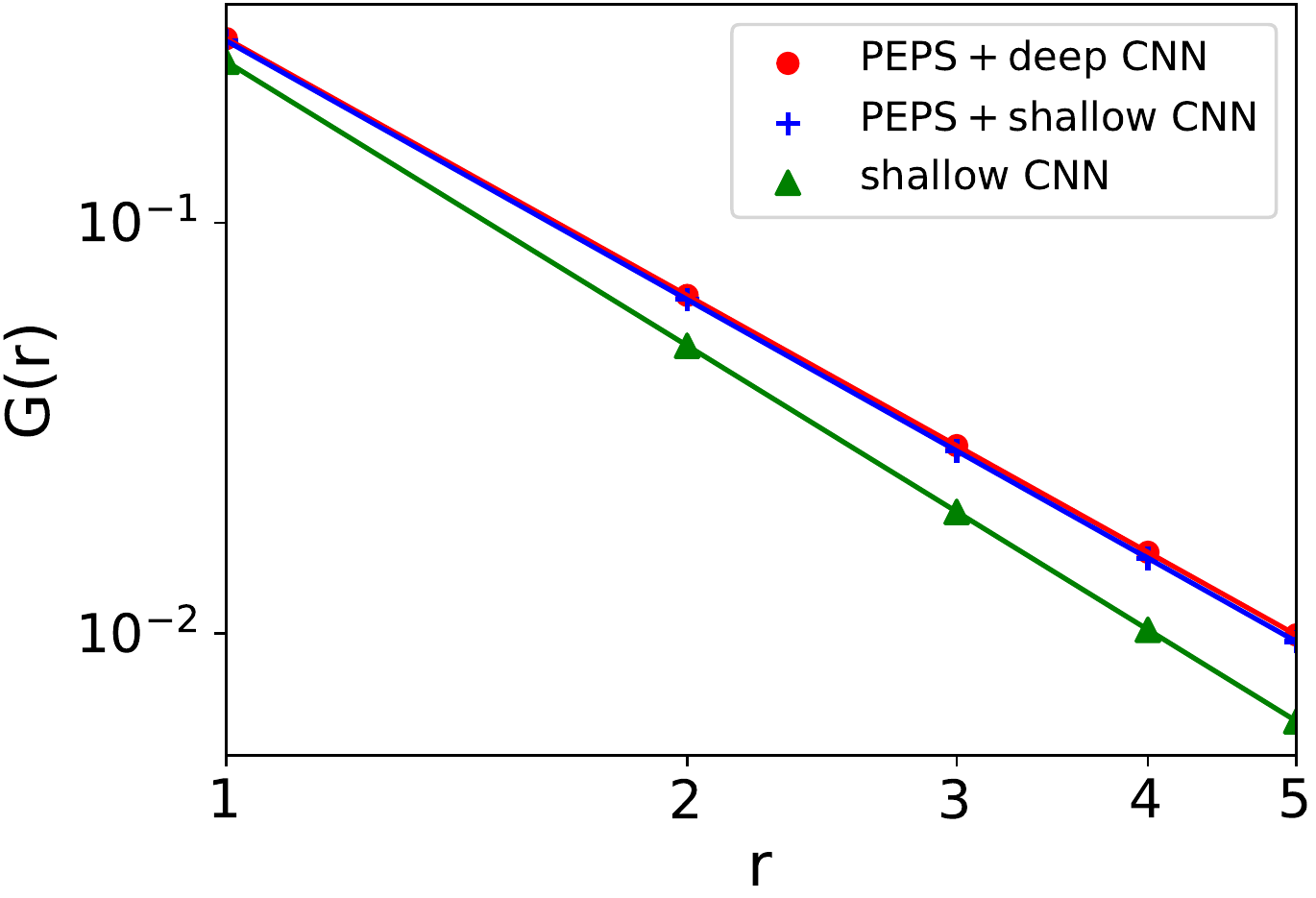}
\caption{Correlation function $G(r)$ with respect to distance $r$ for different CNN wave functions, in the case of $J_2$=0.5.}
%All the wave functions are enforced with rotational symmetry, and the sample number for the spin correlation is $10^6$.}
\label{Fig:correlation}
\end{figure}

We first give the results of the shallow CNN used in literature\cite{CNNj1j2_1} without a prior sign structure but with enforcing rotational symmetry, for $J_2$=0 and $J_2$=0.5. The convolution filter number $M_1$=128 and $M_2$=1 and the side length of the filter $K$=9, are used. The number of variational parameters of this NN is 11009. The network is optimized by the Stochastic Reconfiguration (SR) method \cite{SR}.
We obtain the energy per site -0.668236 for $J_2$=0 and -0.482986 for $J_2$=0.5, which are listed in Table I.

We then benchmark the shallow CNN multiplied by a prior sign structure provided by PEPS.
We reduce the number of filters for the CNN structure to $M_1$=90 and $M_2$=1 with filter size $K$=9.
The total variational parameters of this structure is reduced to 7741 for the CNN.
Since the computational scaling of a PEPS of PBC is very high,
we use a PEPS with open boundary condition, despite
the physical system has PBC. The PEPS bond dimension is taken to be $D$=1, which
allows the PEPS can be efficiently contracted.
\begin{table*}[]
\caption{Comparison between the ground energies of the spin-1/2 $J_1$-$J_2$ model on a 10$\times$10 square lattice,
achieved by this work (PEPS+deep CNN) and other NN based wave function methods in the literatures.
The GWF+RBM stands for the RBM-enhanced Gutzwiller projected fermionic wave functions, whereas CNN1 and CNN2 are two convolutional NNs taken from Ref.\cite{CNNj1j2_2} and Ref.\cite{CNNj1j2_3} respectively.}
\begin{tabular}{c|ccccc}
\hline\hline
          & GWF+RBM\cite{ansatz3} & \ \ \ CNN1\cite{CNNj1j2_2}\ \ \      & \ \ \ CNN2\cite{CNNj1j2_3}\ \ \          & \ \ \ Best\cite{AFM_QMC,J1J2_best}\ \ \  & This work  \\ \hline
$J_2$=0   & -0.67111   & -0.67135    & -0.671275      & -0.671549   & -0.671330  \\
$J_2$=0.5 & -0.49575   & -0.49516    & -0.494757      & -0.49755    & -0.495502  \\
\hline
\end{tabular}
\label{Tab:best_comparison}
\end{table*}

To obtain the sign structure, we first optimize the PEPS wave function alone (without CNN)
for $J_2$=0 using the imaginary time-evolution method with the so-called simple update scheme\cite{PEPS_4}.
We exam the signs of the wave-function on massive random spin configurations, and find that the sign rule
provided by PEPS is consistent with MPSR.
After we obtain the PEPS for the sign structure,
the CNN wave function is then optimized with the SR method, by fixing the PEPS sign rules
as a precondition factor.
For $J_2$=0, the ground state energy is calculated to be $E$=-0.670718, which is significantly better than that without the precondition.

We find that if we use the PEPS optimized by SU under $J_2$=0.5 for the sign rule, even with the bond dimension as large as $D=4$, the energy converges very
slowly when further optimize the CNN. Since it has been shown in Ref.~\onlinecite{CNNj1j2_2,CNNj1j2_3} that
the MPSR is also a good precondition factor for $J_2$=0.5,
we use the PEPS sign rule obtained from $J_2$=0 for the $J_2$=0.5 case.
The converged energy per site is -0.492335 for $J_2$=0.5, also significantly better than
that without the precondition.

We now benchmark the PEPS and deep CNN hybrid wave function for the $J_1$-$J_2$ model.
We use the same PEPS sign rule as used for the above shallow CNN.
The energy convergence with respect to SR steps is depicted in Fig.~\ref{Fig:Precisions}.
%When performing SR to optimize PEPS+deep CNN,
We start with random parameters for the deep CNN and the wave functions are optimized
under fixed PEPS.
We first optimize the network without enforcing rotational symmetry.
For $J_2$=0, the energy converges quickly with respect to the SR steps as shown in Fig.~\ref{Fig:Precisions}(a).
After 170 steps, the energy converges to -0.670697. We then enforce the rotational symmetry, the energy further reduces to -0.671330.

The CNN is much more difficult to optimize in the case of $J_2$=0.5.
As shown in Fig.~\ref{Fig:Precisions}(b), it takes about 2880 SR epochs to converge without enforcing the rotational symmetry, and the energy per site is -0.493587. By enforcing the rotational symmetry, the energy further reduces to -0.495502.

We compare the energies obtained by several state-of-art NN methods in Table.~\ref{Tab:best_comparison}.
The GWF+RBM wave function \cite{ansatz3} is constructed as the product of a Gutzwiller-projected fermionic state
and a complex-valued RBM. CNN1 and CNN2 are the two convolutional networks presented in Ref.\cite{CNNj1j2_2}
and Ref.~\cite{CNNj1j2_3}, respectively.

The ground state energy for the non-frustrated model, i.e., $J_2$=0
can be obtained to very high accuracy by the stochastic series expansion method~\cite{AFM_QMC}.
The best ground state energy in the literature for $J_2$=0.5 is obtained
by a Gutzwiller projected fermionic wave function improved
by the Lanczos iterations\cite{J1J2_best}.

For $J_2$=0, our energy is only about 2$\times$10$^{-4}$ higher than the best result.
Compared to other NN based variational wave function methods,
our result is only about 2$\times$10$^{-5}$ higher
than the one given by CNN1~\cite{CNNj1j2_2}, but are better than GWF+RBM and CNN2.

For $J_2$=0.5, GWF+RBM gives the best results among the NN methods,
but still about 1.8$\times$10$^{-4}$ higher than the best result in
the literature.\cite{J1J2_best}. Our energy is about 2.5$\times$10$^{-4}$
higher than the GWF+RBM method, but lower than the other two CNN methods.

The CNN1~\cite{CNNj1j2_2} has 3838 complex numbers, a totally 7676 variational parameters much more
than 3531 parameters used in our deep CNN.
Since the sign structures used in the works are almost identical,
our results suggested that the maxpooling is an efficient way to capture the long-range entanglement, and the product of neurons is an efficient way to generate the signs of the wave-functions.

We also evaluate the antiferromagetic order parameter
\begin{equation}
	S^2(\textbf{q})=\frac{1}{N(N+2)}\sum_{i,j}\langle\textbf{s}_i\cdot\textbf{s}_j \rangle e^{i\textbf{q}\cdot(\textbf{r}_i-\textbf{r}_j)},
	\label{order_parameter}
\end{equation}
where $\textbf{q}=(\pi,\pi)$ and $\textbf{q}=(\pi,0)$ corresponds to the structure factor of Neel order and stripe order. The spin correlations $\langle\textbf{s}_i\cdot\textbf{s}_j \rangle$ are calculated using $10^6$ samples.
For $J_2$=0, we have $S^2(\pi,\pi)$=0.15665 and $S^2(\pi,0)$=0.00496, whereas for $J_2$=0.5,
we have  $S^2(\pi,\pi)$=0.05880 and $S^2(\pi,0)$=0.00633.
These results are consistent with those given in Ref.~\onlinecite{CNNj1j2_2} and Ref.~\onlinecite{CNNj1j2_3}.

In Fig.~\ref{Fig:correlation}, we compare the spin correlation functions obtained by different CNN structures, in the case of $J_2$=0.5,
%The correlation function is calculated based on the spin correlations under $10^6$ samples:
\begin{equation}
	C(r)=\frac{1}{2L^2}\sum_{i,j}(\langle \textbf{s}_{i,j}\cdot\textbf{s}_{i+r,j}\rangle+\langle \textbf{s}_{i,j}\cdot\textbf{s}_{i,j+r}\rangle)\, .
	\label{correlation}
\end{equation}
The results show that the correlation functions obey power-lay decay.
We fit $C(r)$ as
 $C(r)=\alpha r^{-\gamma}+c_0$, we plot the correlation functions $G(r)=\alpha r^{-\gamma}$, i.e., the constants have been subtracted.
We obtain $\gamma$=2.30, 2.09 and 2.08 for the shallow CNN, PEPS+shallow CNN and PEPS+deep CNN, respectively.
We see that the correlation function without the PEPS sign structure decays much faster than those with the sign structures,
which may cause the errors in the ground state energy.

We mark that this is only a proof-of-principles work where the PEPS sign structures are fixed
beforehand, and therefore, the ground state energies we obtain is in-principle
only the upper bound of the energies that can be achieved by this ansatz.
Further optimization of the PEPS sign structure along with the
optimization of CNN is necessary, especially
for the models whose sign rule can not be reached beforehand.
However, because the sign function is a jump function, making the optimization
the sign structure very challenging.
We leave this problem for future investigations.

\section{Summary}
%\textit{Conclusions}--

We proposed a variational ansatz for the quantum many-particle wave functions which
combines two state-of-art techniques, i.e., PEPS and deep CNN.
In this ansatz, CNN represents the wave function amplitudes, whereas PEPS provides the sharp changing
sign structures which are difficult to be presented by the deep CNN structure.
We test this ansatz on the two-dimensional spin-1/2 $J_1$-$J_2$ model.
We demonstrate that even a PEPS of small bond dimension can present the sign rule very well,
and the hybrid structure can achieve very accurate ground state energy that competitive to
results of other state-of-the-art neural networks.
Further optimization of the PEPS sign structure along the optimization of CNN is a promising
routine to improve the results and to solve more general models.
%We leave this problem for future investigations.

\section{Acknowledgement:}

X.L thanks for the inspiring discussion with Li Chen. Implementation of SR was discussed with Giuseppe Carleo.
X. L. acknowledges support from the Beijing Outstanding Young Scientist Program, Ministry of Science and Technology Grant No. 2016YFA0301600.
L. H. acknowledges support from the Chinese National Science Foundation Grants No. 11774327.
S. D. acknowledges the support from China Postdoctoral Science Foundation funded Grant No. 2018M632529.
The deep CNN is built by PyTorch.
The calculations were done partly on the HPC of USTC,
as well as the computational resources in Prof. Hui Zhai's group.

\end{document}